\documentclass{article}

\usepackage{PRIMEarxiv}

\usepackage[utf8]{inputenc} % allow utf-8 input
\usepackage[T1]{fontenc}    % use 8-bit T1 fonts
\usepackage{hyperref}       % hyperlinks
\usepackage{url}  
\usepackage{float}% simple URL typesetting
\usepackage{booktabs}       % professional-quality tables
\usepackage{amsfonts}       % blackboard math symbols
\usepackage{nicefrac}       % compact symbols for 1/2, etc.
\usepackage{microtype}      % microtypography
\usepackage{lipsum}
\usepackage{fancyhdr}       % header
\usepackage{graphicx}       % graphics
\graphicspath{{media/}}     % organize your images and other figures under media/ folder

\title{The use of new technologies to support Public Administration. \\
\large Sentiment analysis and the case of the app 'IO'}

\author{
  Vincenzo Miracula\\
  Dipartimento di Fisica e Astronomia “E. Majorana” \\
  Università di Catania \\
  Catania\\
  \texttt{vincenzo.miracula@phd.unict.it} \\
   \And
  Antonio Picone \\
  Dipartimento di Fisica e Astronomia “E. Majorana” \\
  Università di Catania \\
  Catania\\
  \texttt{antonio.picone@phd.unict.it} \\
 }

\begin{document}
\maketitle

\begin{abstract}

App IO is an app developed for the Italian PA. It is definitely useful for citizens to interact with the PA and to get services that were not digitized yet. Nevertheless, it was not perceived in a good way by the citizens and it has been criticized. As we wanted to find the root that caused all these bad reviews we scraped feedback from mobile app stores using custom-coded automated tools and - after that - we trained two machine learning models to perform both sentiment analysis and emotion detection to understand what caused the bad reviews.

\end{abstract}


\section{Introduction}

Since 2005, there has been an increasing development of digitization within the public administration that sees the introduction of the use of technology as a privileged tool in the management of administrative activities. The main objective is to promote digitization in administrations in order to achieve greater efficiency in their activities in internal relations, between different administrations, and between the latter and private individuals. The entry of artificial intelligence into public action, however, needs to be accompanied by an adequate regulatory framework to guarantee the rights of those administered.

The notion of digital transformation has gained significant attention in the literature\cite{vukvsic2018preliminary}. Although approaches to the definition of digital transformation vary\cite{reis2018digital}, most authors suggest that digital transformation involves the use of ICT technology to create fundamentally new capabilities in business, public administration\cite{lankshear2008digital} and people's lives\cite{westerman2011digital}. The theory emphasizes the importance of digitization to optimize the public value of government services for citizens\cite{bannister2014ict} as well as to increase the efficiency of government functions by implementing lean government models\cite{janssen2013lean}. Digital technologies stimulate citizen participation\cite{luna2017opportunities} and support customer involvement in the co-production and co-creation of public value\cite{cordella2018icts}.
The digital transformation thus has a significant impact not only on the accessibility and quality of public services but also on the way other functions of public administration are carried out\cite{dobrolyubova2021measuring} (like policy development, regulation, and enforcement, etc.).

A strong emphasis in the current phase is placed on government solutions as a platform\cite{o2011government} and, consequently, on data-driven governance. In the academic literature, there are also other approaches to the staging of digital transformation in public administration. Janowski\cite{janowski2015digital}, for instance, suggests that digital transformation is not a one-off event or project with a clear start and end date, but rather an ongoing process involving changes both in internal processes and procedures and in the way the public administration communicates with its main beneficiaries. 

It is assumed that the results of this process will be significant, but so far no uniform approaches have been implemented to measure these results. Although digital transformation is an evolving process, the measures used to evaluate this process should also evolve according to the stage of the digital transformation. For instance, in the early stages of e-government, indicators such as the ratio of e-services provided online are considered relevant. 

In the more mature stages of digital transformation, the effects are different: smart government leads to the replacement of official portals by automated interactions and, thus, results in the reduction of the types of public services\cite{dobrolyubova2021measuring}. This contribution intends to present new tools for measuring the results of digital transformation through the use of sentiment analysis and social network analysis.  

\section{The use of the IO app in public administration}

As of 18 April 2020, the IO app is downloadable from the online stores for Android and Apple. The app, conceived and developed by the Digital Transformation Team, was still in Beta version in April 2020, with a limited number of services available, but making it downloadable was an important first step to start testing it on a large scale. The IO app, in fact, is the platform designed to allow all citizens to have a new and single telematic access point to the services, information, and communications of the public administration, thus enabling them to use national and local public services from their smartphones in a simple, modern and secure way.

Two years after its release on online stores, there have been many developments, mainly related to the fact that the app was chosen as a tool to access some of the services activated in the months of the Covid-19 pandemic, such as the holiday bonus, the 'Cashback' program and the green-pass. The regulatory basis of the IO project is Art. 64 bis of the new Digital Administration Code\cite{cad2022} (CAD), which provides a single point of telematic access to public administration services and highlights a fundamental change in the relationship between citizens and PA, centered on three key aspects: simplicity, speed, and transparency. The ultimate goal is to bring the citizen closer to the administration, making simple mechanisms that often still prove to be complex and cumbersome. The IO app is, therefore, the tool designed to concretely enable Digital Citizenship, providing citizens with a direct connection (through their smartphone) with PA services and communications, a single access point for the provision and use of public services.

As can be understood, for now, it is always the citizen who has to express his or her willingness to interact with the Public Administration through the IO app. By using the 'IO' app, the citizen agrees to interface with the Public administration he or she needs to get in touch with according to the provisions of the 2019 Three-Year Plan for Information Technology in Public Administration.

In the document 'Strategy for Technological Innovation and the Digitisation of the Country'\cite{PNI2025} (also called, Plan 2025) presented on 17 December 2019 by Minister Pisano, the IO app is included among the top 20 Actions to Transform the Country.  In fact, page 15 of the document states that "Italy, at least digitally, should be one big municipality that treats all its citizens equally", and again, "IO, is the public services app that transforms the relationship between citizens and Public Administration, putting people at the center and erasing complexity: a single interface to access all public services directly from your smartphone after identifying yourself with your digital identity." 

\section{Methodology}

In order to extract data for the purposes of our analysis, we decided to extract user reviews from the two main stores for mobile devices\cite{hemphill2017intellectual}: AppStore for Apple iOS users and Play Store for Android users. To date, both account for 95\% of total downloads, so they constitute a representative sample of the thinking of the users of such applications in the Italian context.

In order to extract the reviews, code was written in Python, so as to automate the extraction of the reviews, not only of the text but also of the author of the review and the stars that were assigned (from 1 to 5). This process is called 'scraping', an English term that literally means 'scraping'. Since there is no official API, which would allow programmatic access, the only alternative for data extraction is scraping. This technique consists of accessing the various tags of the HTML source code in order to extract the desired fields. This approach certainly requires higher skills than the use of an API and a further cleaning of the data, sometimes soiled by the presence of HTML tags that must be removed, in order to obtain text in its classic form.

The dataset, consisting of the cleaned textual reviews and the metadata mentioned, was saved in serialized JSON (JavaScript Object Notation) format, for easy reading by both a human being and a computer, as further analysis of sentiment and emotions followed, obtained through the use of machine learning techniques applied to Natural Language Processing. In particular, BERT (Bidirectional Encoder Representations from Transformers) was used, a type of neural network, based on Transformers, devised by Google\cite{von2017advances}, thanks to whose attention mechanism, each word within the same sentence is related both to the word that follows it, and to the previous one, allowing the entire context of a sentence to be maintained.

We therefore, on the basis of an existing model for Italian, carried out what is called fine-tuning, i.e. starting from a general model, not specific to any task, the model was trained with our data to perform and perform better in the two tasks specific to our objective. In this way, we obtained a sentiment analysis model capable of discerning sentences into two categories: positive and negative; furthermore, as already mentioned, a model for emotion detection was also realized. There has long been debate as to which emotions are worth training our model on it is necessary to have a psychological background on the matter and the literature on the subject has proved to be of fundamental importance. Paul Ekman\cite{ekman1992argument} is an American psychologist who came up with the universal theory of emotions, according to which there are seven primary emotions: fear, anger, joy, sadness, contempt, disgust, and surprise; but these have been reformulated and amalgamated to become six in its latest official version, namely: fear, anger, joy, sadness, disgust, and surprise. However, according to a recent study\cite{jack2014dynamic}, it is proposed that human beings have four basic emotions: fear, anger, joy, and sadness. Over the course of time, other authors have also proposed the above emotions as the four basic emotions\cite{gu2015differentiation},\cite{wang2016neuromodulation}. In light of these studies, our model follows in the footsteps and ideas of the most recent studies on the subject, thus managing to categorize input sentences into the four categories of emotions mentioned above.

\section{Network analysis }

Social media platforms have become key channels for communication and information dissemination, making it increasingly important to understand how information spreads within these networks. In general, in social network analysis, nodes are people and ties are all the social connections between them - for example, friendship, marital/family ties, or financial ties. 

Social network analysis (SNA) is a field of study involving the use of statistical and mathematical techniques to analyze and understand relationships and patterns within a network of individuals or organizations. It is often used to identify key actors and understand the dynamics of social, professional, and communication networks. The objective of social network analysis is to understand a community by mapping the relationships that connect it as a network and then trying to identify key individuals and groups within the network and/or associations between individuals.

Twitter is an excellent platform to observe these behaviors. For this reason, we decided to collect tweets to try to understand what are the dynamics that lead to the formation of communities on Twitter, mapping the relationships that connect people as a network and then trying to bring out the key individuals, and groups within the network and associations between individuals.

In an SNA, a network is typically represented as a graph, with nodes representing individual actors and links representing relationships between them (e.g. friendship, collaboration). Several metrics can be used to analyze these networks, including measures of centrality (e.g. degree) measures of community structure (e.g. modularity), and measures of network evolution (e.g. preferential attachment). 

SNA has a wide range of applications, including studying the spread of information and ideas, identifying influencers, understanding the structure and dynamics of social networks, and predicting the formation of new relationships. It is used in fields such as sociology, psychology, anthropology, communication, and information technology and has also been applied to the analysis of networks in business, politics, and public health.

\section{Results}

%Social network analysis seeks to understand networks and their participants and has two main focuses: the actors and the relationships between them in a specific social context. 
Social networks represent an emerging challenging sector where the natural language expressions of people can be easily reported through short but meaningful text messages. Key information that can be grasped from social environments relates to the polarity of text messages.

The first result of our research is related to the citizen's perception of the IO app. We then extracted 62986 reviews from the various stores and through the use of Natural Language Processing\cite{chowdhary2020natural} techniques such as sentiment analysis and emotion detection, we have that 75.91\% of the reviews have negative sentiment and 60.5\% negative emotions (34.3\% sadness, 26.2\% anger). 

The second result relates to SNA. From our analysis, the SNA presents 5 different and distinct communities calculated according to Louvain's algorithm\cite{blondel2008fast}. The community detection thus constructed shows moderately connected communities with a modularity value of 0.6. We also note that the graph presents itself according to a Barabasi-Albert\cite{albert2002statistical} with 62986 nodes with an average degree (i.e. the average number of links) of 1.37 and a density of 0.573(\ref{tab1}). 

\begin{table}(H)
    \caption{Network results}
    \centering
    \begin{tabular}{|l|c|}
    \hline
      Statistics & Value \\
      Number of nodes & 62986 \\
      Mean degree & 1.37 \\
      Density & 0.573 \\
      Modularity & 0.6\\
      \hline
    \end{tabular} 
    \label{tab1}
\end{table}

\begin{figure}
	\centering
		\includegraphics[width=\linewidth]{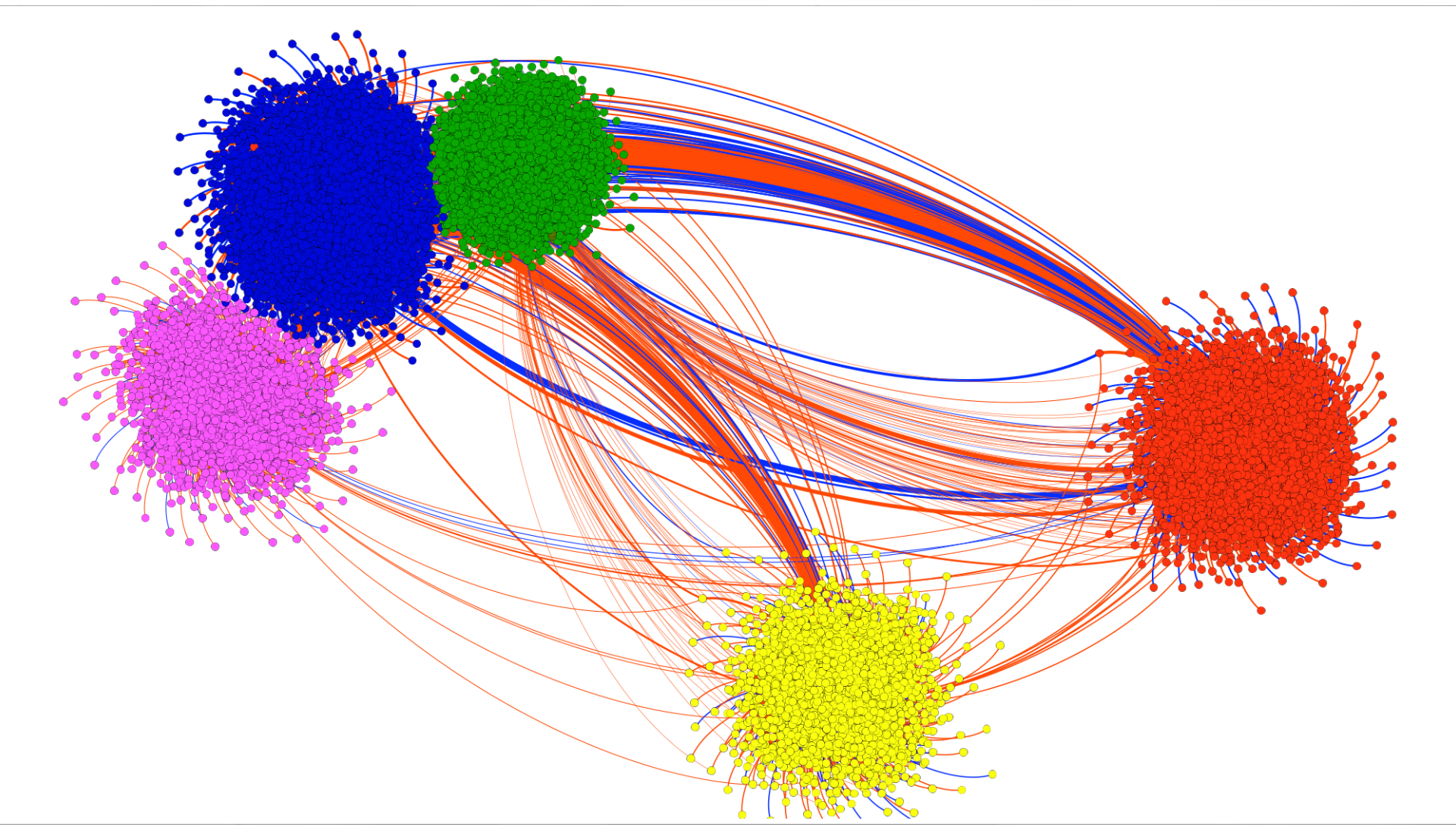}
	\caption{network analysis IO App.}
	\label{fig:1}
\end{figure}

These results, as shown in Fig. \ref{fig:1} lead to the hypothesis that individuals struggle to change their opinion and sentiment toward the use of the IO app (and by extension towards the digitization path undertaken by the PA). 

\section{Conclusion}

This paper brings attention to how computational social science and network science can both explain the complex dynamics of controversial and challenging topics. The digital ecosystem not only evolves social network communication but also provides the social researcher with useful data to explain social-complex dynamics.

Using natural language processing techniques, such as sentiment analysis and emotion detection, we proposed a new way to measure the impact of digital transformation and how the resulting analysis can be applied to provide legislators with meta-evaluation tools, in this case starting with how the app is perceived by citizens.

We noticed that many of the comments found within the dataset we created are not related to the app itself, a large number of them are negative towards the government and the PA itself. The comments are therefore just a proxy for the citizen to criticize policy choices.

\bibliographystyle{unsrt}  
\bibliography{bibliography.bib}

\end{document}